\documentclass{article}
\usepackage{geometry} 
\usepackage{amsmath}
\usepackage{amssymb}
\usepackage{accents}
\usepackage{latexsym}
\usepackage{authblk}
\usepackage{hyperref}
\usepackage{graphicx}
\usepackage{bmpsize}
\usepackage[square,numbers]{natbib}  
\usepackage{color} 
\hypersetup{
setpagesize=false,
 bookmarksnumbered=true,%
 bookmarksopen=true,%
 colorlinks=true,%
 linkcolor=blue,
 citecolor=red,
}
\makeatletter
 
  \@addtoreset{equation}{section}
\makeatother

\title{Scalar Mode Quadrupole Radiation from Astronomical Sources in $F(R)$ Modified Gravity}

\author[1,2,3]{Tomohiro Inagaki }
\author[3]{Masahiko Taniguchi}

\affil[1]{Information Media Center, Hiroshima University, Higashi-Hiroshima, 739-8521, Japan}
\affil[2]{Core of Research for the Energetic Universe, Hiroshima University, Higashi-Hiroshima, 739-8526, Japan}
\affil[3]{Graduate School of Advanced Science and Engineering, Hiroshima University, Higashi-Hiroshima, 739-8526, Japan }

\date {} 
\begin{document}
\maketitle

\begin{abstract}
We investigate the scalar mode quadrupole radiation of gravitational waves in $F(R)$ modified gravity.
In $F(R)$ gravity a massive scalar mode appears in the gravitational waves. 
 We find explicit expressions for the quadrupole radiation and  the energy current of the scalar mode in general $F(R)$ gravity models.
 We consider a binary star and a bouncing star as astronomical sources of the gravitational waves and calculate the quadrupole radiation of the scalar and tensor modes. 
 The scalar mode radiates under spherically symmetric conditions, but the tensor modes do not.
 The scalar mode mass is estimated for some typical energy scales.
We show a possibility to detect the scalar mode in the future gravitational waves observation.
\end{abstract}

\newpage
\section{Introduction}
$F(R)$ gravity is a modified gravity theory in which the Einstein-Hilbert action, $R$, is replaced by a general function of $R$. 
It has been introduced as non-linear generalization of Einstein's theory\cite{Buchdahl:1983zz}.
One of the major applications of this idea has been made on the construction of cosmological models with an accelerating expansion\cite{Starobinsky:1979ty}.
Numerous models have been proposed to explain the early and late-time accelerating expansion of the universe, as a review, see, for example,
\cite{Nojiri:2010wj,Clifton:2011jh,Nojiri:2017ncd}.

There is potential to test the models of $F(R)$ gravity by looking at current observations attributed to the expansion of the universe, for example, type Ia supernovae\cite{Riess:1998cb,Aghanim:2018eyx}, CMB fluctuations\cite{Samtleben:2007zz,Planck:2018jri} and BAO\cite{SDSS:2005xqv,Bassett:2009mm}.
Evidence of accelerating expansion alone is not sufficient, and other procedures to test the model of F(R) gravity are being explored.
In fact, several studies have been done on the verification of F(R) gravity through the equation of state inside neutron stars\cite{Kase:2019dqc,Dohi:2020bfs} and its contribution to the solar system\cite{Guo:2013fda}.
In this paper, we focus on the possibility of testing the models of $F(R)$ gravity in gravitational waves.

The first direct detection of gravitational waves(GWs) from a binary black hole was succeeded in 2015 by LIGO\cite{LIGOScientific:2016aoc}.
This is a new clue in examining the theory of gravity.
In consequence, the observed gravitational waves were consistent with the predictions of general relativity(GR).
It shows that GR can be adapted to strong gravity.
However, GWs may directly reveal the existence of phenomena beyond GR.
Exploring extra modes of GWs has already been done\cite{LIGOScientific:2018czr}.
Expectations are growing for the development of future GWs detectors such as KAGRA\cite{Aso:2013eba}, LIGO-India\cite{Unnikrishnan:2013qwa}, LISA\cite{Danzmann:2003tv} and DECIGO\cite{Kawamura:2011zz}.

One of the characteristics of F(R) gravity is that an extra degree of freedom appears in GWs\cite{Alves:2009eg,Myung:2016zdl,Gong:2017bru,Moretti:2019yhs,Gogoi:2020ypn,Chowdhury:2021wrq}.
The extra degree of freedom propagates as a scalar mode of GWs.
The scalar mode of $F(R)$ gravity has a non-vanishing mass depending on $F(R)$ modification\cite{Yang:2011cp,Sharif:2017ahw,Capozziello:2017vdi,Capozziello:2019klx,Katsuragawa:2019uto,Gogoi:2019zaz,Kalita:2021zjg}.
Thus, $F(R)$ gravity can be constrained through the scalar mode mass\cite{DeLaurentis:2011tp,Lambiase:2020vul,Vainio:2016qas,Nojiri:2017hai,Lee:2017dox,Narang:2022jkv}.

Here, we investigate the scalar mode propagation in $F(R)$ gravity in more detail.
First of all, we try to solve the wave equation with a source. Applying the procedure in Ref.\cite{{Berry:2011pb}} to a general $F(R)$ gravity, the wave equation can be divided into tensor and scalar modes.
Then we study the gravitational waves propagation from gravitational sources.
We solve the wave equation for the scalar mode and evaluate the quadrupole radiation.
The scalar mode radiation is considered from two typical sources, a binary star and a bouncing star that shrinks in size and bounces back.
The amplitude of the scalar mode is suppressed by the mass correction.
We calculate the suppression compared with the tensor modes.
Then we estimate the possibility to detect the scalar mode in future gravitational wave observations.
We also evaluate the delay of the massive scalar mode from the first signal according to the propagation speed.

This paper is organized as follows. Sec. 2 describes the basic formulation for the tensor and scalar modes of the gravitational wave. We give expressions for the quadrupole radiation and energy current. In Sec. 3 we evaluate the scalar mode radiation from a binary star and a bouncing star and discuss the possibility to detect the scalar mode. Finally, we give some concluding remarks.

\section{Basic formulation}
\subsection{Wave equation}
$F(R)$ gravity is motivated by an exploration of cosmic accelerating expansion such as the inflation and dark energy by extension of the Ricci scalar, $R$ to a general form, $F(R)$ in the action. 
It is expected that $F(R)$ gravity induces phenomena beyond GR. We focus on the possibility to test the model of $F(R)$ gravity through gravitational wave propagation. 

We start from the $F(R)$ gravity action,
\begin{align}\label{eqs:FR action}
S=\int d^4x \sqrt{-g}\frac{1}{16\pi G}F(R) + S_{matter}.
\end{align}
where $G$ denotes the gravitational constant.
The equation of motion is driven by varying the action (\ref{eqs:FR action}) with respect to the metric tensor,
\begin{align} \label{eqs:EoM of FR}
\mathcal{G}_{\mu\nu}=8\pi G {T_{\mu\nu}},
\end{align}
where we introduce the modified Einstein tensor, $\mathcal{G}_{\mu\nu}$, defined by 
\begin{align}\label{eqs:modified Einstein tensor}
\mathcal{G}_{\mu\nu}\equiv F'(R)R_{\mu\nu}-\frac{1}{2}g_{\mu\nu}F(R)+(g_{\mu\nu}\Box-\nabla_\mu\nabla_\nu)F' ,
\end{align}
and ${T_{\mu\nu}}$ is the energy-momentum tensor derived from the matter action, $S_{matter}$.
 To find the gravitational wave equation, the metric perturbation is employed in Eq.(\ref{eqs:modified Einstein tensor}).
We consider the perturbation of the metric tensor around a flat Minkowski background, $\eta_{\mu\nu}$,
\begin{align}\label{eqs:perturb metric}
g_{\mu\nu}=\eta_{\mu\nu}+h_{\mu\nu}.
\end{align}
The perturbation of $F(R)$ and $F'(R)$ around the background curvature $\tilde{R}$ is given by
\begin{align}\label{eqs:f(r) action perturbation} 
F(R) &=F(\tilde{R})+F^{\prime}(\tilde{R}) \delta R,
\\ 
F^{\prime}(R) &=F^{\prime}(\tilde{R})+F^{\prime \prime}(\tilde{R}) \delta R.
\end{align}
The scalar mode of GWs is identified with
\begin{align}
\frac{F^{\prime \prime}(\tilde{R})}{F'(R)} \delta R =\Phi.
\end{align}
It should be noted that the curvature, $\tilde{R}$, vanishes in the flat Minkowski background.

The gravitational wave equation (\ref{eqs:EoM of FR}) contains a mixture of tensor and scalar modes.
We extend the prescription separating these two modes in Ref.\cite{Berry:2011pb} to a general $F(R)$ gravity.
To find a wave equation for the physical degrees of freedom we introduce, $\bar{h}_{\mu\nu}$,
\begin{align}\label{eqs:bar h munu}
 \bar{h}_{\mu\nu}&=h_{\mu\nu}+(b\Phi-\frac{1}{2}h)\eta_{\mu\nu},
\end{align}
 and impose the following gauge conditions,
\begin{align}\label{eqs:gauge condition}
\nabla^\nu \bar{h}_{\mu\nu}=0.
\end{align}
In these conditions
the lowest order of Ricci tensor and scalar are expressed as
\begin{align}\label{eqs:perturbation of Ricci}
R^{(1)}_{\mu\nu}
&=-\frac{1}{2}\left[\square\left( \bar{h}_{\mu\nu}-\frac{\bar{h}}{2}\eta_{\mu\nu}\right)+b(\eta_{\mu\nu}\square\Phi+2\partial_{\mu}\partial_{\nu}\Phi)\right],
\\ \label{eqs:perturbation of Ricci2}
R^{(1)} &=\frac{1}{2}\square \bar{h}-3b\square\Phi.
\end{align}
We set $T_{\mu\nu}=0$ and derive the perturbed equation of motion from Eq.(\ref{eqs:EoM of FR}),
\begin{align}\label{eqs:perturbation of EOM}
R^{(1)}_{\mu\nu}-\frac{1}{2}{\eta}_{\mu\nu}R^{(1)} +\left[{\eta}_{\mu\nu}\square - \partial_\mu\partial_\nu \right]\Phi =0.
\end{align}
The perturbed equation is divided into the tensor and scalar parts,
\begin{align}\label{eqs:perturbation of EOM2}
-\frac{1}{2}\square \bar{h}_{\mu\nu}+(b+1)\left[{\eta}_{\mu\nu}\square - \partial_\mu\partial_\nu \right]\Phi =0.
\end{align}

To eliminate the scalar part from Eq.(\ref{eqs:perturbation of EOM2}), we set $b=-1$. 
Then Eq.(\ref{eqs:perturbation of EOM2}) reduces to
\begin{align}\label{eqs:wave eq of tensor}
\square \bar{h}_{\mu\nu}=0.
\end{align}
The wave equation of the tensor mode is now successfully extracted and equivalent to the one in GR.
Thus the GWs propagation of tensor mode in $F(R)$ gravity is nothing changes from GR. 
The transverse-traceless gauge conditions can be also imposed to $\bar{h}$ as in GR,
\begin{align}\label{eqs:tensor mode relation}
\eta^{\mu\nu}\bar{h}_{\mu\nu}=0, \quad \bar{h}_{0i}=0.
\end{align}
On the other hand, the scalar mode equation can be obtained by tracing Eq.(\ref{eqs:perturbation of EOM}).
\begin{align}\label{eqs:Wave eq of scalar}
\left[\square-m_{F(R)}^2 \right]\Phi=0,
\end{align}
where the mass squared in Eq.(\ref{eqs:Wave eq of scalar}) is expressed as,
\begin{align}\label{eqs:mass in FR}
m_{F(R)}^2= \frac{1}{3}\frac{F'({0})}{F''({0})}.
\end{align}
The existence of scalar mode is attributed to the $F(R)$ modification.
In other words, physics beyond general relativity emerges. 
This is because the F(R) modified gravity has an extra degree of freedom\cite{Myung:2016zdl,Gong:2017bru,Moretti:2019yhs}
The mass in Eq.(\ref{eqs:mass in FR}) depends on the function $F(R)$\cite{Yang:2011cp,Sharif:2017ahw,Capozziello:2017vdi,Katsuragawa:2019uto,Gogoi:2019zaz,Kalita:2021zjg}.
The scalar mode shows the verifiability of modified gravity theory through the GWs detections\cite{Vainio:2016qas,Nojiri:2017hai}.

\subsection{Tensor modes}
To consider the phenomena of GWs, 
the energy-momentum tensor is induced in the wave equation for tensor mode (\ref{eqs:wave eq of tensor}) as a source of GWs,
\begin{align}\label{eqs:tensor mode}
\square \bar{h}_{\mu\nu}=8\pi \tilde{G} T_{\mu\nu}.
\end{align}
where we redefine the gravitational constant as $\tilde{G}=G/F'$.
We find the radiation of gravitational waves from the solution of this equation.

It is more convenient to employ the Fourier representation of $\bar{h}$,
\begin{align}\label{eqs:tensor mode eq}
\bar{h}_{\mu\nu}(\textbf{x},t)=\frac{1}{\sqrt{2\pi}}\int dk^0 \bar{h}_{\mu\nu}(\textbf{x},k^0)e^{-ik^0t}.
\end{align}
Green's function is defined by the solution of $\square G({\bf x})=\delta({\bf x})$. 
By using the Fourier representation of Green's function, the tensor mode solution of Eq.(\ref{eqs:tensor mode}) is found to be
\begin{align}\label{eqs:Green function of tensor}
\bar{h}_{\mu\nu}(\textbf{x},k^0)=-16\pi \tilde{G}\int d^3{\bf x}G({\bf x-x'},k_0) T_{\mu\nu}({\bf x'},k_0).
\end{align}
As is well-known, the Fourier representation of Green's function is given by
\begin{align}\label{eqs:Green funciton x k0}
G({\bf x},k_0)=\frac{1}{\sqrt{2\pi}^3}\int\frac{1}{{\sqrt{2\pi}}^3}\frac{1}{-{\bf k}^2+k_0^2} e^{i{\bf k\cdot x}}d^3{\bf k}
=-\frac{1}{4\pi|{\bf x}|}e^{ik_0 |{\bf x}|}.  
\end{align}
Substituting Eq.(\ref{eqs:Green funciton x k0}) into Eqs.(\ref{eqs:Green function of tensor}) and (\ref{eqs:tensor mode eq}), we obtain the retarded solution of the tensor mode,
\begin{align}
\label{eqs:tensor mode sol}
\bar{h}_{\mu\nu}(\textbf{x},t)={4\tilde{G}}\int d^3{\bf x'}\frac{ T_{\mu\nu}({\bf x'},t-|{\bf x-x'}|)}{|{\bf x-x'}|}.
\end{align}
This solution shows that GWs emitted from the source travel at the speed of light.
The only difference in tensor mode between GR and F(R) gravity is the gravitational constant. If $\tilde{G}$ is regarded as the observed constant, no difference appears.

\subsection{Scalar mode}
For the scalar mode propagation from gravitational sources, the trace of the energy-momentum tensor is introduced in the wave equation (\ref{eqs:Wave eq of scalar}),
\begin{align}
[\square-m^2]\Phi={8\pi\tilde{G}} T.
\end{align}
Green's function for scalar mode is defined as the solution of $[\square-m^2 ]\mathcal{G}({\bf x})=\delta({\bf x})$. 
The difference from Green's function in the tensor mode is the non-vanishing mass.
After the integral with respect to the wave vector, we obtain the Fourier representation of the green's function,
\begin{align*}
\mathcal{G}({\bf x},k_0)=\frac{1}{\sqrt{2\pi}^3}\int\frac{1}{{\sqrt{2\pi}}^3}\frac{1}{-{\bf k}^2-m^2+k_0^2} e^{i{\bf k\cdot x}}d^3{\bf k}
=-\frac{1}{4\pi|{\bf x}|}e^{i\sqrt{k_0^2-m^2} |{\bf x}|}.
\end{align*}
The Fourier representation of the scalar mode is given by
\begin{align}\label{eqs:Green function of scalar}
\Phi(\textbf{x},k^0)=-16\pi\tilde{G}\int d^3{\bf x}\mathcal{G}({\bf x-x'},k_0) T({\bf x'},k_0).
\end{align}
By the inverse Fourier transformation the scalar mode is represented as
\begin{align}\nonumber
\Phi(\textbf{x},t)&=-16\pi\tilde{G}\int dk^0\int d^3{\bf x'}\mathcal{G}({\bf x-x'},k_0) T({\bf x'},k_0)e^{-ik^0t} 
\\ \label{eqs:phi with green}
&={4\tilde{G}}\int d^4{x'}{ T({\bf x'},t')}\mathcal{G}(|{\bf x-x'}|,t-t').
\end{align}

To simplify the expression we set $t-t'=\Delta t$.
Then, Green's function, $\mathcal{G}(|{\bf x-x'}|,\Delta t)$, is rewritten as
\begin{align}
\mathcal{G}(|{\bf x-x'}|,\Delta t)=\frac{1}{2\pi |{\bf x-x'}|}\int_{-\infty}^{\infty}e^{i(\sqrt{{k^0}^2-m^2}|{\bf x-x'}|-k_0\Delta t)}dk^0.
\end{align}
The Green's function is represented by the Bessel function according to Ref.\cite{morse1953methods,Naf:2011za},
\begin{align}
\label{eqs:green for scalar}
\mathcal{G}(|{\bf x-x'}|,\Delta t)
=\frac{\delta(\Delta t-|{\bf x-x'}|)}{|{\bf x-x'}|}-\frac{m}{\sqrt{\Delta t^2-|{\bf x-x'}|^2}}J_1(m\sqrt{\Delta t^2-|{\bf x-x'}|^2})\theta(\Delta t- |{\bf x-x'}|).
\end{align}
Substituting Eq.(\ref{eqs:green for scalar}) to Eq.(\ref{eqs:phi with green}), the retarded solution of the scalar mode is found to be
\begin{align}
\Phi(\textbf{x},t)
&={4\tilde{G}}\int d^3{\bf x'}\left[\frac{ T({\bf x'},t-|{\bf x-x'}|)}{|{\bf x-x'}|}\right.
\nonumber
\\
\label{eqs:phi by bessel}
& \left.-\int_{-\infty}^{t-|{\bf x-x'}|} dt'\frac{m}{\sqrt{\Delta t^2-|{\bf x-x'}|^2}}J_1(m\sqrt{\Delta t^2-|{\bf x-x'}|^2}) T({\bf x'},t')\right].
\end{align}
We introduce the time-dependent parameters, $t_p=t-|{\bf x-x'}|$, $t_f=t+|{\bf x-x'}|$ and write $\tau=\sqrt{(t'-t_p)(t'-t_f)}$.
Then the second term on the right-hand side in Eq.(\ref{eqs:phi by bessel}) is rewritten as
\begin{align}\label{eqs:Bessel with tau}
\int_{V} d^3{\bf x'}\int^{\infty}_{0} md\tau \frac{J_1(m\tau)}{m\sqrt{\tau^2+|{\bf x-x'}|^2}} T({\bf x'},t-\sqrt{\tau^2+|{\bf x-x'}|^2}).
\end{align}
We transform the integral variable $\tau$ to $\zeta$ with $m\tau=m|{\bf x-x'}|\sinh{\zeta}$.
Eq.(\ref{eqs:Bessel with tau}) is simplified to
\begin{align}\label{eqs:scalar expression right-hand}
\int_{V} d^3{\bf x'}\int^{\infty}_{0} d\zeta {J_1(m|{\bf x-x'}|\sinh{\zeta})} T({\bf x'},t-\cosh{\zeta}|{\bf x-x'}|).
\end{align}
Therefore the scalar mode is found be
\begin{align} 
\Phi({\bf x},t)
\label{eqs:scalar expression}
&={4\tilde{G}}\int d^3{\bf x'} \int_0^{\infty}d\zeta \left[\frac{\delta(\zeta)}{|{\bf x-x'}|}-mJ_1(m|{\bf x-x'}|\sinh\zeta) \right] T({\bf x'},t-\cosh\zeta |{\bf x-x'}|).
\end{align}
It should be noted that $\cosh{\zeta}$ takes the value from $1$ to $\infty$ for the interval of the integration, $\zeta:0\to\infty$.
We regard $c_s\equiv 1/\cosh\zeta$ as the velocity of scalar mode propagation.
For $\zeta=0$ the velocity is equal to the speed of light.
At the limit $\zeta\to\infty$ the scalar mode does not propagate, i.e. $c_s =0$.

\subsection{Quadrupole radiation}
Here we focus on GW radiation whose source is sufficiently far away from the observer and the gravitational source is non-relativistic.
In this case, the GW radiation is
generated by the quadrupole and higher moments of the energy and momentum distributions.

First, we consider the tensor mode solution.
For the observer, $|{\bf x-x'}|\sim |{\bf x}|\equiv r$, Eq.(\ref{eqs:tensor mode sol}) is written as
\begin{align} \label{eqs:tensor mode GR source}
\bar{h}^{ij}({\bf x},t)&\sim\frac{4\tilde{G}}{r}\int T^{ij}({\bf x'},t-r)d^3{\bf x'}.
\end{align}
at the leading order.
From the conservation law, $\partial_\mu  T^{\mu\nu}=0$,
we obtain
\begin{align*}
\partial_{\mu}\partial_{\nu}x^ix^jT^{\mu\nu}(x)
=
2 T^{ij}(x).
\end{align*}
The three-dimensional spatial integration of this equation gives
\begin{align}
\nonumber
\int  2 T^{ij}(x)d^3{\bf x}
&=
\int d^3{\bf x}
\left[
\partial_{m}\partial_{l}x^ix^jT^{ml}
+\partial_{0}\partial_{0}x^ix^j T^{00}(x)+2\partial_{k}\partial_{0}x^ix^j(T^{k0}(x)+T^{0k}(x))
\right]
\\
\label{eqs:2nd derivative energy momentum}
&=
\partial_{0}\partial_{0}\int d^3{\bf x}
x^ix^j T^{00}(x).
\end{align}
From the first line to the second line in this equation, we drop the surface terms. 
Substituting Eq.(\ref{eqs:2nd derivative energy momentum}) into Eq.(\ref{eqs:tensor mode GR source}), the tensor mode solution is represented as the 2nd derivative of the quadrupole moment, $I^{ij}$, 
\begin{align} \nonumber
\bar{h}^{ij}({\bf x},t)&
\sim\frac{2\tilde{G}}{r}\partial_{0}\partial_{0}\int  T^{00}({\bf x'},t-r){x^i}'{x^j}'d^3{\bf x'}
\\ \label{eqs:quadrupole moment for tensor}
&=\frac{2\tilde{G}}{r}\frac{d^2I^{ij}}{d^2t}.
\end{align}
It should be noted that the projection operators are necessary to describe the polarization of the tensor modes.

Next, we move to the scalar mode solution (\ref{eqs:scalar expression}).
The change of variable, $w=mr\sinh\zeta$, makes the integral of the Bessel function easier to compute \cite{Moriguchi:1987muh},
\begin{align}
\int_0^{\infty}d\zeta mJ_1(mr\sinh\zeta)=\int_0^{\infty}dw \frac{mJ_1(w)}{\sqrt{w^2+(mr)^2}} 
=m I_{\frac{1}{2}}\left(\frac{mr}{2}\right)K_{\frac{1}{2}}\left(\frac{mr}{2}\right),
\end{align}
where $I_{\frac{1}{2}}$ and $K_{\frac{1}{2}}$ denote the modified Bessel functions and satisfy,
\begin{align}
&I_{\frac{1}{2}}(z)=\sqrt{\frac{2}{\pi z}}\sinh z,\,
K_{\frac{1}{2}}(z)=\sqrt{\frac{\pi}{2 z}}e^{-z} .
\end{align}
For the distant observer, it is assumed that the velocity of scalar mode propagation is almost constant and the energy-momentum tensor is independent on the value of $\cosh\zeta$.
In other words, assuming that the energy-momentum tensor does not depend on the velocity or the velocity changes a little.
Then Eq.(\ref{eqs:scalar expression}) becomes
\begin{align} \nonumber
\Phi&={4\tilde{G}}\int d^3{\bf x'} \int_0^{\infty}d\zeta \left[\frac{\delta(\zeta)}{r}-mJ_1(mr\sinh\zeta) \right] T({\bf x'},t-\frac{r}{c_s})
\\ \nonumber
&\sim{4\tilde{G}}\int d^3{\bf x'}\left[\frac{1}{r}-\frac{1-e^{-mr}}{r} \right] T({\bf x'},t-\frac{r}{c_s})
\\
&=\frac{4\tilde{G}e^{-mr}}{r}\int d^3{\bf x'} T({\bf x'},t-\frac{r}{c_s}).
\end{align}
The tracing of the energy-momentum tensor can be divided into 00 and spatial parts.
$T_{00}$ gives the mass density and $T_{ij}$ is related to the 2nd derivative of the quadrupole moment as can be seen from the tensor mode analogy,
\begin{align}
\Phi&= \frac{4\tilde{G}e^{-mr}}{r}\int Td^3{\bf x'}
=\frac{4\tilde{G}e^{-mr}}{r}\int ({ T^0}_0+{ T^i}_i)d^3{\bf x'}
\\ \label{eqs:scalar quadrupole}
&=\frac{4\tilde{G}e^{-mr}}{r}M+\frac{2\tilde{G}e^{-mr}}{r}\frac{d^2I}{d^2t}.
\end{align}
The scalar mode has a Yukawa-like potential depending on the total mass and the trace of the quadrupole moment, $I$.
This property is due to the fact that the scalar mode is massive, which has been obtained in other studies\cite{Katsuragawa:2019uto,Kalita:2021zjg}.
The additional quadrupole radiation part enables us to understand dynamical phenomena.

\subsection{Energy current}
Following the procedure developed in \cite{Berry:2011pb}, we calculate the effective energy-momentum tensor for a general form of $\Phi$.
The gravitational radiations carry energy and then act as a source of gravitational fields.
In order to introduce the background metric arising from gravitational waves themselves we have to consider the perturbation around a curved background metric, $\gamma_{\mu\nu}$.
\begin{align}\label{eqs:dev metric around BG}
g_{\mu\nu}=\gamma_{\mu\nu}+h_{\mu\nu}.
\end{align}

The perturbation of the modified Einstein tensor can be described as,
\begin{align*}
\mathcal{G}_{\mu\nu}=\mathcal{G}^{\rm B}_{\mu\nu}+\mathcal{G}^{(1)}_{\mu\nu}+\mathcal{G}^{(2)}_{\mu\nu},
\end{align*}
where the number in the upper indices denotes the order of the expansion and $\mathcal{G}^{\rm B}_{\mu\nu}$ is the modified Einstein tensor for the background.
For the GWs the 1st order term vanishes, $\mathcal{G}^{(1)}_{\mu\nu}=0$, from the wave equation.
Then the background satisfies
\begin{align}\label{eqs:background of GWs}
\mathcal{G}^{\rm B}_{\mu\nu}=-\mathcal{G}^{(2)}_{\mu\nu}.
\end{align}
Later, we will average over several wavelengths, $<\mathcal{G}^{(2)}_{\mu\nu}>$, assuming that the background is on a large scale compared with the wavelengths of GWs.
Eq.(\ref{eqs:background of GWs}) means that the background metric $\gamma_{\mu\nu}$ is $\mathcal{O}(h^2)$. 
So the background is decomposed
\begin{align}
\gamma_{\mu\nu}=\eta_{\mu\nu}+j_{\mu\nu},
\end{align}  
where $j_{\mu\nu}$ is the order $\mathcal{O}(h^2)$.
The curvature tensor of the background is also $\mathcal{O}(h^2)$.
Then the 2nd-order perturbation of the Ricci tensor is given by
\begin{align}\nonumber
R^{(2)}_{\mu\nu}=&\frac{1}{4}\nabla_{\mu}h^{\alpha\beta}\nabla_{\nu}h_{\alpha\beta}+\frac{1}{2}h^{\alpha\beta}(\nabla_{\mu}\nabla_{\nu}h_{\alpha\beta}+\nabla_{\alpha}\nabla_{\beta}h_{\mu\nu}-\nabla_{\alpha}\nabla_{\nu}h_{\mu\beta}-\nabla_{\alpha}\nabla_{\mu}h_{\beta\nu})
\\ \label{eqs:2nd perturbation of Ricci tensor}
&-\frac{1}{2}(\nabla_{\beta}h^{\alpha\beta}-\frac{1}{2}\nabla^\alpha h)(\nabla_{\nu}h_{\mu\alpha}+\nabla_{\mu}h_{\alpha\nu}-\nabla_{\alpha}h_{\mu\nu})+\frac{1}{2}\nabla^{\beta}{h^\alpha}_\nu(\nabla_{\beta}h_{\mu\alpha}-\nabla_{\alpha}h_{\mu\beta}).
\end{align}
From Eq.(\ref{eqs:bar h munu}) with $b=-1$ it is  expressed by $\bar{h}$ and $\Phi$ as,
\begin{align}\label{eqs:2nd perturbation of Ricci tensor2}
R^{(2)}_{\mu\nu}=\frac{1}{4}\nabla_{\mu}\bar{h}^{\alpha\beta}\nabla_{\nu}\bar{h}_{\alpha\beta}+\frac{1}{2}\bar{h}^{\alpha\beta}\nabla_{\mu}\nabla_{\nu}\bar{h}_{\alpha\beta}+\frac{3}{2}\nabla_{\mu}\Phi\nabla_{\nu}\Phi+\Phi\nabla_{\mu}\nabla_{\nu}\Phi+\frac{1}{2}\gamma_{\mu\nu}\Phi\square\Phi.
\end{align}
Thus the 2nd-order modified Einstein tensor (\ref{eqs:modified Einstein tensor}) is found to be
\begin{align} \nonumber
{\mathcal{G}^{(2)}}_{\mu\nu}=&F'\left[R^{(2)}_{\mu\nu}-\frac{1}{2}\gamma_{\mu\nu}R^{(2)}-\frac{1}{2}h_{\mu\nu}R^{(1)}\right]+F''\left[R^{(1)}R^{(1)}_{\mu\nu}-\frac{1}{4}\gamma_{\mu\nu}{R^{(1)}}^2\right]
\\  \nonumber
&+\gamma_{\mu\nu}\square(F''R^{(2)})-\gamma_{\mu\nu}h^{\alpha\beta}F'\partial_{\alpha}\partial_{\beta}\Phi+h_{\mu\nu}\square(F''R^{(1)})-\gamma_{\mu\nu}\gamma^{\alpha\beta}{\Gamma^\rho}_{\alpha\beta}^{(1)}F'\partial_{\rho}\Phi
\\ \label{eqs:perturb 2nd einstein tensor}
&-\partial_{\mu}\partial_{\nu}(F''R^{(2)})+{\Gamma^\rho}_{\mu\nu}^{(1)}F'\partial_{\rho}\Phi,
\end{align}
where the perturbation of the connection is 
\begin{align*}
{\Gamma^\rho}_{\mu\nu}^{(1)}&=\frac{1}{2}\gamma^{\rho\lambda}(\partial_{\mu}h_{\lambda\nu}+\partial_{\nu}h_{\mu\lambda}-\partial_{\lambda}h_{\mu\nu})
\\
&=\frac{1}{2}\gamma^{\rho\lambda}(\partial_{\mu}\bar{h}_{\lambda\nu}+\partial_{\nu}\bar{h}_{\mu\lambda}-\partial_{\lambda}\bar{h}_{\mu\nu})-\frac{1}{2}\gamma^{\rho\lambda}(\gamma_{\lambda\nu}\partial_{\mu}\Phi+\gamma_{\mu\lambda}\partial_{\nu}\Phi-\gamma_{\mu\nu}\partial_{\lambda}\Phi).
\end{align*}

On a large scale background curvature, the terms that remain after averaging over several wavelengths are
\begin{align}\label{eqs:ave 2nd perturb Ricci tensor}
\left<R^{(2)}_{\mu\nu}\right>=\left<-\frac{1}{4}\partial_{\mu}\bar{h}^{\alpha\beta}\partial_{\nu}\bar{h}_{\alpha\beta}+\frac{1}{2}\partial_{\mu}\Phi\partial_{\nu}\Phi+\frac{1}{2}\gamma_{\mu\nu}\Phi\square\Phi\right>,
\end{align}
and
\begin{align}\label{eqs:ave 2nd perturb Ricci scalar}
\left<R^{(2)}\right>&=\left<\gamma^{\mu\nu}R^{(2)}_{\mu\nu}-h^{\mu\nu}R^{(1)}_{\mu\nu}\right>
=\left<\frac{9}{2}\Phi\square\Phi\right>.
\end{align}
The averages over several wavelengths for Eqs.(\ref{eqs:perturbation of Ricci}) and (\ref{eqs:perturbation of Ricci2}) are
\begin{align*}
\left<R^{(1)}_{\mu\nu}\right>&=\frac{1}{2}\left<\gamma_{\mu\nu}\square\Phi+2\partial_{\mu}\partial_{\nu}\Phi\right>,
\\
\left<R^{(1)}\right>&=\left<3\square\Phi\right>,
\end{align*}
where we use the wave equation Eq.(\ref{eqs:wave eq of tensor}) so we take $\square\bar{h}_{\mu\nu}=0$.
Then we get
\begin{align*}
\left<R^{(1)}R^{(1)}_{\mu\nu}\right>&=\left<3\square\Phi\partial_{\mu}\partial_{\nu}\Phi+\frac{3}{2}\gamma_{\mu\nu}(\square\Phi)^2\right>.
\end{align*}
Thus, we obtain the average of the 2nd order perturbation of modified Einstein tensor (\ref{eqs:perturb 2nd einstein tensor}), 
\begin{align}\label{eqs:ave modified einstein tensor}
\left<\mathcal{G}^{(2)}_{\mu\nu}\right>=F'\left<-\frac{1}{4}\partial_{\mu}\bar{h}^{\alpha\beta}\partial_{\nu}\bar{h}_{\alpha\beta}-\frac{3}{2}\partial_{\mu}\Phi\partial_{\nu}\Phi\right>.
\end{align}

The effective energy-momentum tensor is defined by
\begin{align}\label{eqs:def GWs tensor}
F'T^G_{\mu\nu}\equiv-\frac{1}{8\pi\tilde{G}}\left<\mathcal{G}^{(2)}_{\mu\nu}\right>.
\end{align}
Substituting Eq.(\ref{eqs:ave modified einstein tensor}) into Eq.(\ref{eqs:def GWs tensor}), we successfully derived the effective energy-momentum tensor including the scalar mode in the general case, $\Phi$.
\begin{align}
T^G_{\mu\nu}=\frac{1}{8\pi\tilde{G}}\left<\frac{1}{4}\partial_{\mu}\bar{h}^{\alpha\beta}\partial_{\nu}\bar{h}_{\alpha\beta}+\frac{3}{2}\partial_{\mu}\Phi\partial_{\nu}\Phi\right>.
\end{align}

By the replacement of $t$ and $r$ the energy current is given by
\begin{align}
\frac{dE_{\rm{GW}}}{dt}=-\int <T^{G}_{0r}(t-r)>r^2d\Omega=\int <T^{G}_{00}(t-r)>r^2d\Omega,
\end{align}
where we take the propagation speed of the scalar mode to almost light speed, $c_s\sim 1$. We will see the validity of this assumption in the later section. 
.
The tensor and scalar modes, $\bar{h}$ and $\Phi$, are described by the 2nd derivative of the quadrupole moment.
For a distant observer the total mass, $M$, is conserved and the time derivative of the first term in Eq.(\ref{eqs:scalar quadrupole}) drops. 
Then the energy current is written in the quadrupole representation,
\begin{align}\label{eqs:energy current}
\frac{dE_{\rm{GW}}}{dt}=\left<\frac{\tilde{G}}{5}\dddot{\mathcal{I}}_{ij}\dddot{\mathcal{I}}^{ij}+12\tilde{G}e^{-2mr}{\dddot{I}}^2\right>,
\end{align}
Eq.(\ref{eqs:energy current}) shows that the scalar mode effects emerge in gravitational radiation in addition to tensor mode.

Blow the gravitational constant $\tilde{G}$ is written as $G$.

\section{Scalar mode quadrupole radiation}
\subsection{Binary star}
At present, GWs from compact binary stars are the most promising source for observations.
We focus on the scalar mode GWs from binary stars.
It is assumed that the binary star rotates on the $xy$ plane and these masses have $m_1,m_2$ and the stellar distance is $L$.
The distances from the center of gravity to each star are given by $(r_1,r_2)=(m_2L/M,m_1L/M)$.

The quadrupole moment is defined by
\begin{align}\label{eqs:Quadrupole}
I_{ij}=\int d^3x' \rho(x'){x'}_i{x'}_j.
\end{align}
The density and position of the binary star are represented as, 
\begin{align*}
&\rho(x)=m_1\delta({\bf x}-{\bf x_1})+m_2\delta({\bf x}-{\bf x_2}),
\\
&{\bf x_1}=(r_1\cos\omega t,r_1\sin\omega t,0),
\\
&{\bf x_2}=(-r_2\cos\omega t,-r_2\sin\omega t,0),\ (\omega=\sqrt{{GM}/{L^3}}).
\end{align*}
After the spatial integration the quadrupole moment of the binary star is derived
\begin{align}\label{eqs:Quadrupole of BS}
I_{ij}=\begin{pmatrix}
L^2\mu\cos^2\omega t &L^2\mu\cos\omega t\sin\omega t & 0\\ L^2\mu\cos\omega t\sin\omega t& L^2\mu\sin^2\omega t & 0 \\
0 & 0 & 0 \\
\end{pmatrix},
\end{align}
where $\mu$ denotes the reduced mass, $\mu\equiv m_1m_2/(m_1+m_2)$.
When we take the typical velocity of the stars $v$ and the distance between Earth and the binary star $r$, the amplitude of tensor mode from a binary star is evaluated by Eq.(\ref{eqs:quadrupole moment for tensor}),
\begin{align}\nonumber
|h_{ij}|&=\frac{4G}{rc^4}\frac{\mu L^2(2\pi f)^2}{r}\sim\frac{4G}{rc^4}\frac{\mu v^2}{r}
\\
&\sim 5\times10^{-23}\left(\frac{100\rm{Mpc}}{r}\right)\left(\frac{\mu}{10M_{\odot}}\right)\left(\frac{v}{0.1c}\right)^2.
\end{align}

On the other hand, the trace of the quadrupole moment (\ref{eqs:Quadrupole of BS}) becomes $I=L^2\mu$.
If the trace of the quadrupole moment does not have time dependence, the scalar mode does not radiate from a binary star.
Since the tensor modes GW carries away the energy of a binary star,
the interstellar distance, $L$ decrease with time~\cite{maggiore2008gravitational} and the scalar mode radiate.
Because of a monotonic time dependence of $L=L_0(1-t/t_{\rm coal})^{1/4}$ the scalar mode may have a chirp signal that does not oscillate.
Thus the amplitude is calculated to be
\begin{align}\nonumber
\Phi&\sim\frac{4G\mu {L_0}^2}{c^4t_{\rm coal}^2r}\sim\frac{2^{18}}{5^2}\frac{G\mu^3}{c^2Mr}\left(\frac{v}{c}\right)
\\
&\sim5\times 10^{-31}\left(\frac{100\rm{Mpc}}{r}\right)\left(\frac{10M_{\odot}}{M}\right)^2\left(\frac{\mu}{10M_{\odot}}\right)^2\left(\frac{v/c}{0.1}\right)^{14}.
\end{align}
where $t_{\rm coal}$ is the time of coalescence.
The strain of the amplitude is extremely small.
It increases over time but is not quite sufficient for observation.
There is little hope to observe the scalar mode GWs from a binary star.
However, we considered only the inspiral phase.
The compact binary coalescence has the phases such as merger and ringdown phases~\cite{Ajith:2007xh}.
It is an interesting topic, although it requires more precise analysis~\cite{Shibata:2005ss,Hotokezaka:2013iia}.
\subsection{Bouncing star}
Let us now study a toy model that we call a bouncing star.
It is far from a real phenomenon such as a supernova explosion. 
However, it does provide some clues about the scalar mode propagation in spherically symmetric gravitational sources. 

We consider a star with the radius $R(t)$ and the density $\rho=M/(\frac{4\pi}{3}R(t)^3)$, where $M$ is the total mass of the star and does not depend on time.
The trace of the quadrupole moment becomes
\begin{align}
I&=\int d^3x' \rho(x'){x'}_i{x'}^i
=4\pi\int_0^{R(t)} dr' \rho{r'}^4
=\frac{3}{5}MR(t)^2.
\end{align}
We assume that the star shrinks and bounces once and write the time evolution of radius as,
\begin{align*}
{R(t)}={R_0}(1-be^{\frac{-(t-t_0)^2}{\tau^2}}).
\end{align*}
The star shrinks to $R_0(1-b)$ and bounces at $t=t_0$.
The bouncing time interval is characterized by $\tau$. 
In this situation, the scalar mode from Eq.(\ref{eqs:scalar quadrupole}) is given by
\begin{align}
\Phi=\frac{48\tilde{G}M{R_0}^2b}{5\tau^2}\frac{e^{-mr}}{r}\left[\left(1-\frac{2(t-t_0)^2}{\tau^2}\right)-b\left(1-\frac{4(t-t_0)^2}{\tau^2}\right)e^{\frac{-(t-t_0)^2}{\tau}}\right]e^{\frac{-(t-t_0)^2}{\tau^2}},
\end{align}
where the static potential is neglected.
The scalar mode is emitted from the star with a spherical symmetry.
This result is interesting because the tensor modes does not radiate from a spherically symmetric objects. 

We estimate the amplitude of the bouncing star.
Applying to the bouncing of the core in a supernova explosion, we find
\begin{align*}
|\Phi|_{\rm typical}
&=\frac{48\tilde{G}M{R_0}^2b}{5\tau^2}\frac{e^{-mr}}{r}
=7.93\times10^{-44}\frac{M{R_0}^2b}{\tau^2}\frac{e^{-mr}}{r}
\\
&=2\times10^{-20}\left(\frac{10\rm{kpc}}{r}\right)\left(\frac{M}{M_{\odot}}\right)\left(\frac{R_0}{6000\rm{km}}\right)^2\left(\frac{1{\rm s}}{\tau}\right)^2,
\end{align*}
where the exponential term is dropped by assuming that the mass is sufficiently small and $b$ approximated to 1.
It shows that a core collapse of a supernova explosion in our galaxy may emit the detectable scalar mode GWs.  
Also, the energy current in this event is estimated from Eq.(\ref{eqs:energy current}),
\begin{align}
\frac{dE_{\rm{GW}}}{dt}\sim \frac{12G}{c^5}\frac{M^2{R_0}^4b}{\tau^8}\sim2\times10^{43}\left(\frac{M}{M_{\odot}}\right)^2\left(\frac{R_0}{6000\rm{km}}\right)^4\left(\frac{1{\rm s}}{\tau}\right)^8 {\rm erg/s}.
\end{align}
The gravitational potential energy released in the supernova collapse is estimated in the order of $10^{53}{\rm erg}$~\cite{Kamiokande-II:1987idp,Bionta:1987qt,Super-Kamiokande:2011lwo}.
The scalar mode GWs cost only $10^{-8}\%$ of total energy emission.
The existence of a scalar mode does not have a significant contribution to supernova explosions and subsequent growth.
\begin{figure}[t]
 \centering
 \includegraphics[keepaspectratio, scale=0.6]{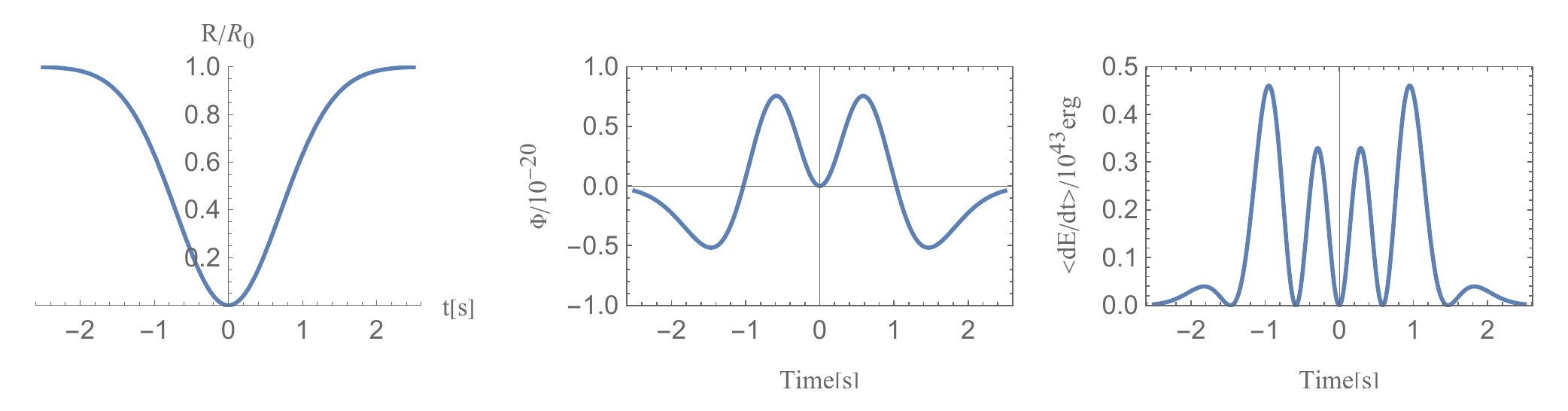}
\caption{Time dependence of radius(left), amplitude(middle), energy current(right) at $b=1$. }
 \label{Fig:SNGWb=1}
\end{figure}
\begin{figure}[t]
 \centering
 \includegraphics[keepaspectratio, scale=0.6]{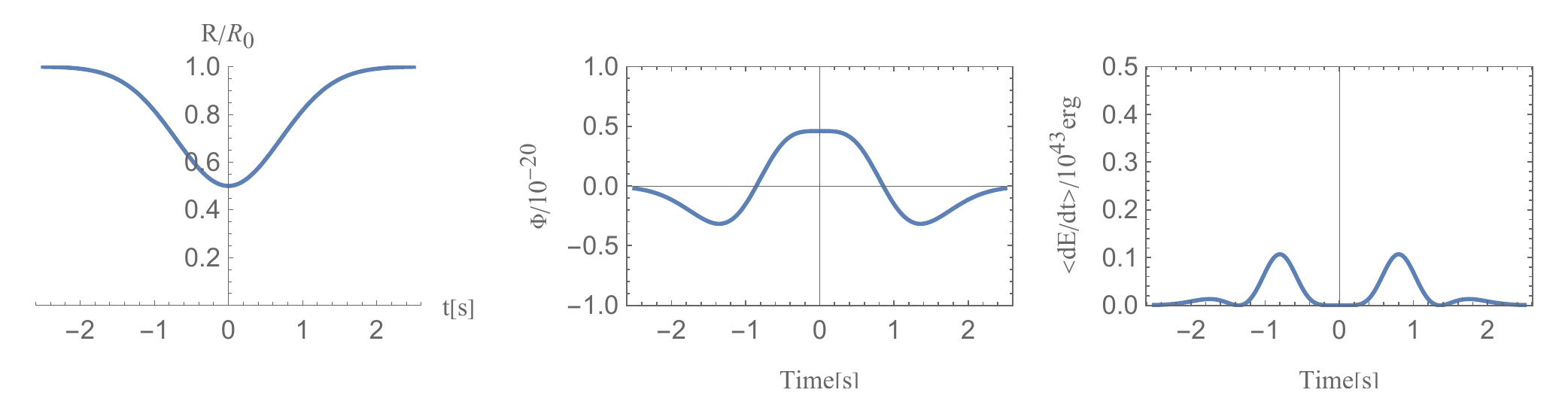}
 \caption{Time dependence of radius(left), amplitude(middle), energy current(right) at $b=0.5$. }
 \label{Fig:SNGWb=05}
\end{figure}

Fig.{\ref{Fig:SNGWb=1}} shows the time dependence of the radius of the bouncing star, amplitude, and energy current of the scalar mode GWs for $b=1$.
We also show the figure in the case of $b=0.5$ in Fig.{\ref{Fig:SNGWb=05}} for comparison.
The parameter $b$ produces a large difference in energy release.
Therefore, a dramatic event, such as the collapse of a star, is necessary to generate detectable scalar mode GWs.

\subsection{Scalar-tensor ratio}
We compare the amplitudes for the tensor and scalar modes.
The scalar-tensor ratio of GWs is defined by
\begin{align}
\mathcal{R}=\frac{|\Phi|}{|h_{ij}|}.
\end{align}
We assume that the quadrupole radiation intensities of the both modes are equivalent, $\ddot{I}=\ddot{I}_{ij}$.
The scalar-tensor ratio only depends on the exponential term in Eq.(\ref{eqs:scalar quadrupole}),
\begin{align}
\mathcal{R}=e^{-mr}.
\end{align}
Below we estimate the ratio in some mass scale of modified gravity.
The scale of mass depends on the modification scale of gravity theory.
For instance, the $R^2$ model\cite{Starobinsky:1979ty}, $F(R)=R+R^2/M^2$ has the scalar mode mass, $m=M/\sqrt{6}$ from Eq.(\ref{eqs:mass in FR}).

For the dark energy scale, $m=10^{-33}{\rm eV}\sim(4200\rm{Mpc})^{-1}$ we obtain
\begin{align}
\mathcal{R}\sim\left(0.999998\right)^{\frac{r}{10\rm{kpc}}},\ \left(0.976\right)^{\frac{r}{100\rm{Mpc}}},\ \left(0.368\right)^{\frac{r}{4200\rm{Mpc}}},
\end{align}
where the distances are assumed as $10\rm{kpc}$ for the scale of the Galaxy, ${100\rm{Mpc}}$ for the scale of galaxy clusters, and ${4200\rm{Mpc}}$ for the scale of primordial gravitational waves. 
Since the attenuation is a few to 60 percent, they are not a major obstacle to observation.
On the other hand, we obtain
\begin{align}
\mathcal{R}\sim\left(10^{-7\times 10^{50}}\right)^{\frac{r}{10\rm{kpc}}},
\end{align}
for the inflation scale, $m=10^{15}{\rm GeV}\sim(2\times 10^{-31}\rm{m})^{-1}$.
In this case, the scalar mode rapidly suppresses.
It seems very difficult to observe the scalar mode with the inflation scale.
Therefore, the scalar mode GWs are interesting observable physical quantities when the typical scale of the modified gravity is at the dark energy scale.

\subsection{Constraints from propagation speed}
We obtain the speed of the scalar mode propagation in Eq.(\ref{eqs:scalar expression}).
Constraints from the propagation speed also help in verification of the scalar mode GWs as well as the scalar-tensor ratio.
The mass constraints are found from the propagation speed in some observation periods.  
The propagation speed for a wave packet is derived as the group velocity, 
\begin{align}
c_s=\frac{\partial\omega}{\partial k}.
\end{align}
The dispersion of the scalar mode is $\omega=\sqrt{k^2+m^2}$.
The propagation speed of the scalar mode becomes
\begin{align}\label{eqs:scalar speed}
c_s=\sqrt{\frac{k^2}{k^2+m^2}}=\sqrt{\frac{\omega^2-m^2}{\omega^2}}.
\end{align}
The tensor modes propagate at light speed and scalar mode does at $c_s$.
From Eq.(\ref{eqs:scalar speed}) the scalar mode mass is estimated as
\begin{align} \nonumber
m&=\omega\sqrt{\left(1-\left(\frac{c_s}{c}\right)^2\right)}
\\ \label{eqs:mass constraints}
&=4.14\times10^{-15}\left(\frac{f}{1\rm{Hz}}\right)\sqrt{1-\left(\frac{c_s}{c}\right)^2}[{\rm eV}/c^2],
\end{align}
where we denote $\omega=2\pi f$ and the light speed $c$ is not omitted.

We write $\Delta t$ as the difference between the arrival time of the tensor and scalar modes at a distance $r$.
It is described as,
\begin{align}
&\Delta t=\frac{r}{c_s}-\frac{r}{c},
\end{align}
Then the ratio of the propagation speeds is estimated as, 
\begin{align}\label{eqs:speed time period}
\frac{c_s}{c}=\frac{1}{1+\frac{c\Delta t}{r}}.
\end{align}
If we detected GWs from inside the Galaxy, $r\sim10\rm{kpc}$, and the maximum delay is a century, the lower bound of $c_s/c$ is determined from Eq.(\ref{eqs:speed time period}),
\begin{align}
\frac{c_s}{c} \geq0.99695.
\end{align}
\begin{table}[tb]
\caption{The upper bound of scalar mode mass $m/\left(\frac{f}{1\rm{Hz}}\right)[{\rm eV}/c^2]$}
  \centering
  \begin{tabular}{lccc}
    \hline
    Period         & 10kpc                  &100Mpc              &4200Mpc \\
    \hline \hline
   a second       &$5.8\times10^{-21}$&$6\times10^{-23}$&$9\times10^{-24}$\\
   a day            &$1.7\times10^{-18}$&$1.7\times10^{-20}$&$2.6\times10^{-21}$\\
   a year           &$3.2\times10^{-17}$&$3.2\times10^{-19}$&$5.0\times10^{-20}$\\
   a century      &$3.2\times10^{-16}$&$3.2\times10^{-18}$&$5.0\times10^{-19}$\\
    \hline
 \end{tabular}
 \label{table:scalar mass}
\end{table}
In a century-long observation, the upper bound of the scalar mass is $3.2\times10^{-16}{\rm eV}/c^2\geq m$ from Eq.(\ref{eqs:mass constraints}).

Table.\ref{table:scalar mass} summarizes the mass constraints for several cases of distance and observation period.
In especially, the scalar mode GW is $\Delta t \sim 3\times10^{-26}{\rm s}$ delay from the tensor modes when the mass is $10^{-33}{\rm eV}/c^2$.
The scalar mode mass in the dark energy scale is difficult to observe because of the tiny delay from the tensor modes.

\section{Conclusion}
We have investigated the quadrupole radiation of GWs in $F(R)$ gravity.
$F(R)$ gravity has an extra degree of freedom in the wave equations beyond GR. 
Thus the scalar mode also radiates in addition to the tensor modes.
The scalar mode has a mass that depends on the $F(R)$ modification.
We have derived the retarded solution in Eq.(\ref{eqs:scalar expression}).
The quadrupole radiation in the scalar mode is represented as a function of the trace of the quadrupole momentum.
It has been shown that the amplitude of the scalar mode is suppressed exponentially.
Also, we have derived the GW energy current including the scalar mode for a general $F(R)$ form in Eq.(\ref{eqs:energy current}).

We have considered the scalar mode radiation from several astronomical sources.
The radiation from binary stars is currently the most successful gravity source for tensor modes but the amplitude is too weak to detect the scalar mode GWs.
However, there is not enough research on the moment of star coalescence and there is room for the observation of the scalar mode radiation.

We have evaluated a simple model of the bouncing star.
The model is not appropriate to adapt to real stars, but it provides some clues to understand the scalar mode radiation.
Spherically symmetric sources emit the scalar mode GWs, not the tensor modes.
Applying the supernova explosions to the bouncing star, we show that the scalar mode radiation from the events inside the Galaxy is possible to detect in future GWs observations.
This phenomenon is expected to be a promising candidate for the detection of scalar mode GWs.
We have calculated the ratio of the amplitude for the scalar and tensor modes and found it proportional to $e^{-mr}$.
If the scalar mode mass is at the dark energy scale, it does not suppress even for a cosmological distance.
On the other hand, it is promptly suppressed at the inflation scale.

The upper bounds on the scalar modes mass have been estimated from the propagation speed constraints in the observation period.
It is much smaller than the inflation scale and larger than the dark energy scale.
In addition to that, in phenomena where tensor modes are hardly radiated, we are able to obtain similar constraints from the photon instead of the massless tensor modes.

We conclude that the verification of $F(R)$ modified gravity using GWs is hopeful for the mode of the current accelerating expansion.
It is difficult to obtain evidence of modification in a high-energy scale such as inflation in the current GW detectors.
The scalar mode can be radiated from a spherically symmetric gravitational source, which is not predicted by GR.
The observation of the scalar mode directly proves the necessity of an extension of GR.

There are other sources of GWs. 
We are interested in GWs from the early universe such as bubble collisions\cite{Hawking:1982ga,Kim:2014ara,Jinno:2017ixd,Jinno:2017fby}.
GWs from high-energy events in the early universe may directly or indirectly influence observations of cosmological phenomena\cite{Ananda:2007xh}.
These phenomena will become important with the next generation of GW observations\cite{Dyadina:2017uue,Capozziello:2008rq,Ricciardone:2016ddg,Pan:2021tpk,Matos:2021qne,Chowdhury:2021wia,Odintsov:2021kup,Odintsov:2022cbm}.
We will continue the work and compare the results in $F(R)$ gravity with other modified gravity theories such as $F(T)$\cite{Cai:2015emx,Cai:2018rzd}, $F(\mathcal{G})$\cite{Nojiri:2005jg,DeFelice:2009wp,Inagaki:2019rhm}, other formalisms, Palatini $F(R)$\cite{Olmo:2011uz} and Cartan $F(R)$\cite{Inagaki:2022blm}.
By examining these in detail, we hope to find the potential of the modified gravity.

\bibliography{ref}
\end{document}